# Bayesian hypothesis testing and hierarchical modelling of ivermectin effectiveness in treating Covid-19

Martin Neil and Norman Fenton
Risk Information and Management Research
School of Electronic Engineering and Computer Science,
Queen Mary University of London

1 October 2021

## Abstract

Ivermectin is an antiparasitic drug that some have claimed is an effective treatment for reducing Covid-19 deaths. To test this claim, two recent peer reviewed papers both conducted a meta-analysis on a similar set of randomized controlled trials data, applying the same classical statistical approach. Although the statistical results were similar, one of the papers (Bryant et al, 2021) concluded that ivermectin was effective for reducing Covid-19 deaths, while the other (Roman et al, 2021) concluded that there was insufficient quality of evidence to support the conclusion Ivermectin was effective. This paper applies a Bayesian approach, to a subset of the same trial data, to test several causal hypotheses linking Covid-19 severity and ivermectin to mortality and produce an alternative analysis to the classical approach. Applying diverse alternative analysis methods which reach the same conclusions should increase overall confidence in the result. We show that there is strong evidence to support a causal link between ivermectin, Covid-19 severity and mortality, and: i) for severe Covid-19 there is a 90.7% probability the risk ratio favours ivermectin; ii) for mild/moderate Covid-19 there is an 84.1% probability the risk ratio favours ivermectin. Also, from the Bayesian meta-analysis for patients with severe Covid-19, the mean probability of death without ivermectin treatment is 22.9%, whilst with the application of ivermectin treatment it is 11.7%. To address concerns expressed about the veracity of some of the studies we evaluate the sensitivity of the conclusions to any single study by removing one study at a time. In the worst case, where (Elgazzar 2020) is removed, the results remain robust, for both severe and mild to moderate Covid-19. The paper also highlights advantages of using Bayesian methods over classical statistical methods for meta-analysis. All studies included in the analysis were prior to data on the delta variant.

A summary version of this article has been published as a letter in the American Journal of Therapeutics.

## 1. Introduction

Recent studies by (Kory, Meduri, Varon, Iglesias, & Marik, 2021) and (Bryant et al., 2021) evaluating the evidence on ivermectin were widely welcomed by those who have argued that this antiparasitic drug is a cheap and effective treatment for Covid-19 infections. The (Bryant et al., 2021) meta-analysis of randomized controlled trials (RCTs) trials concluded:

> *"Moderate-certainty evidence finds that large reductions in COVID-19 deaths are possible using ivermectin. Using ivermectin early in the clinical course may reduce numbers progressing to severe disease. The apparent safety and low cost suggest that ivermectin is likely to have a significant impact on the SARS-CoV-2 pandemic globally."*

These conclusions contrast with those in (Popp et al., 2021) and, in particular, to those in (Roman et al., 2021)[1] which conducted a similar meta-analysis to (Bryant et al., 2021) using a subset of the trials. They concluded:

> *"In comparison to SOC or placebo, IVM did not reduce all-cause mortality, length of stay or viral clearance in RCTs in COVID-19 patients with mostly mild disease. IVM did not have effect on AEs or SAEs. IVM is not a viable option to treat COVID-19 patients."*

A similar, negative conclusion was made earlier by the World Health Organization (WHO) (WHO (World Health organization), 2021). Despite their own meta-analysis of the 7 best randomized control trials available at that time showing that ivermectin reduces mortality by 81%, they concluded:

> *"a recommendation against the use ivermectin in patients with COVID-19 of any severity, except in the context of a clinical trial".*

The conclusions in (Popp et al., 2021), (WHO (World Health organization), 2021) and (Bryant et al., 2021) are not, however, based on the results of the statistical analysis of the data, which were very similar to those of (Bryant et al., 2021) (in fact, statistically, the WHO analysis provides overwhelming support for the effectiveness of ivermectin with a risk ratio 0.19 and 95% confidence interval (0.09, 0.38)). Instead, as claimed in (Fordham & Lawrie, 2021), these conclusions are based on a somewhat vague and possibly biased subjective assessment of the quality of the trials themselves, and erroneously conclude "no effect" from what was merely weaker evidence of a positive effect. The WHO's report recommendation on ivermectin is also inconsistent (based on the evidence presented) with this recommendation in the same report:

> *"a strong recommendation for systemic corticosteroids in patients with severe and critical COVID-19"*

Unlike the previous studies, this paper applies a Bayesian approach (Gelman et al., 2013; Sutton & Abrams, 2001), to a subset of the same trial data, to test several causal hypotheses linking Covid-19 severity and ivermectin to mortality. Applying diverse alternative analysis methods, which reach the same conclusions, should increase overall confidence in the result. We do not consider the many subjective/medical criteria used to determine the "quality" of the studies.

A Bayesian approach also brings with it several advantages over the classical statistical approaches applied to this trials data thus far. Firstly, it allows the evaluation of competing causal hypotheses; so here we test whether Covid-19 mortality is independent of Covid-19 severity, treatment or both treatment and severity. Also, given that a causal link can be established, a Bayesian approach can explicitly evaluate the strength of impact of that causal

---

[1] Note that (Crawford, 2021) has highlighted errors in the data and the analysis carried out by (Roman et al., 2021)

link on mortality. These advantages can be obtained within a Bayesian meta-analysis framework using a hierarchical model which can also take account of 'zero' frequency results which are not estimable in the classical statistical framework. Finally, the Bayesian approach to confidence intervals leads to the ability to directly interpret confidence intervals in a way that does not rely on notions of repeated trials, making them easier to understand.

To address recent widely publicised concerns about the veracity of some of the studies (notably that of (Elgazzar et al., 2020) we also show results from conducting a 'remove one study at a time' sensitivity analysis.

A summary version of this article has been published as a letter in the American Journal of Therapeutics (Neil & Fenton, 2021).

## 2. Trials Data Used

The trials data[2] analysed in our meta-analysis is summarised in Table 1 and is based on (Bryant et al., 2021) Figure 4 (which also provides the full references to the individual studies). In contrast to (Bryant et al., 2021), we have made the following necessary changes:

- We have excluded the study by (Niaee et al., 2021) in our analysis because the severe Covid-19 patients were not separated from the mild/moderate Covid-19 patients in the trial.
- The ivermectin group of the (Lopez-Medina 2021) trial reported zero deaths in 200 patients. However, (Bryant et al., 2021) analysed potential protocol violations and included in the ivermectin group 75 patients removed by Lopez-Medina 2021 but included in their supplementary materials[3]. In our analysis we have used zero in 200 patients (as did Roman et al. 2021)

Also note that the ivermectin and control groups of the (Ravkirti et al., 2021) study have 55 and 57 patients respectively not 57 and 58 as stated in (Roman et al., 2021).

|  | Ivermectin | | Control | |
|---|---|---|---|---|
|  | Total | Deaths | Total | Deaths |
| **Severe Covid-19 trials** | | | | |
| Elgazzar 2020 | 100 | 2 | 100 | 20 |
| Fonseca 2021 | 52 | 12 | 115 | 25 |
| Gonzalez 2021 | 36 | 5 | 37 | 6 |
| Hashim 2020 | 11 | 0 | 22 | 6 |
| Okumus 2021 | 36 | 6 | 30 | 9 |
| | | | | |
| **Mild/moderate Covid-19 trials** | | | | |
| Ahmed 2020 | 45 | 0 | 23 | 0 |
| Babalola 2020 | 42 | 0 | 20 | 0 |
| Chaccour 2020 | 12 | 0 | 12 | 0 |
| Elgazzar 2020 | 100 | 0 | 100 | 4 |
| Hashim 2020 | 48 | 0 | 48 | 0 |
| Lopez-Medina 2021 | 200 | 0 | 198 | 1 |
| Mahmud 2020 | 183 | 0 | 180 | 3 |
| Mohan 2021 | 100 | 0 | 52 | 0 |
| Petkov 2021 | 50 | 0 | 50 | 0 |
| Ravikirti 2021 | 55 | 0 | 57 | 4 |
| Rezai 2020 | 35 | 1 | 34 | 0 |
| ***Total*** | ***1105*** | ***26*** | ***1078*** | ***78*** |

Table 1: Trial data used in this Bayesian Meta-analysis

---

[2] The full citations reference for the studies are provided in (Bryant et al 2021) and are not repeated here.
[3] Lawrie et al private correspondence.

## 3. The Bayesian Meta-analysis

The Bayesian meta-analysis approach has several stages involving learning from data, determining which causal hypotheses best explain this data, selecting the 'best' hypothesis and then using this to estimate its impact. The stages are linked as follows:

A. Learn the mortality probability distribution from relevant trials for each hypothesis of concern using a hierarchical Beta-Binomial model.
B. For each causal hypothesis use the model in stage A to learn the mortality probability distributions relevant to that causal hypothesis.
C. For each causal hypothesis use the learnt probability distributions from stage B to predict the observed data and calculate the likelihood of observing the data.
D. For all causal hypotheses compute the posterior probability of each hypothesis given the likelihood of observing the data under that hypothesis and select the most likely causal hypothesis that explains the data.
E. Estimate the magnitude of impact of the relevant variables, under the selected 'best' hypothesis, on mortality.

In the Bayesian approach the data are fixed and the model parameters are unknown and are estimated from the fixed data. We apply a Beta-Binomial Bayesian learning model for Stage A in our meta-analysis, a method commonly used in Bayesian statistical learning. For each trial, $i$, we use a Binomial model to learn the distribution of the mortality probability, $p_i$, within that trial from the trial data ($n_i$ $patients, x_i$ $deaths$). Next, we learn the underlying mortality probability distribution, $p$, which explains the trial probability distributions, $p_i$ and for this purpose the beta distribution is the natural choice. The Beta distribution has two parameters $\alpha$ and $\beta$ which make it sufficiently flexible to accurately model a wide range of (not necessarily symmetric) distribution shapes. In any Bayesian approach we must provide 'prior' probabilities for all the parameters to be learnt. As specified in the Appendix the parameters are all given an "ignorant prior" distribution meaning that any possible value is equally likely. This expresses our ignorance about all mortality probability parameters. It is important to note that alternative models and priors (such as a model which includes an exponential prior for sample size) produce very similar results to those reported here. In other words, the analysis is not sensitive to the particular (reasonable) prior assumptions made.

Model computation uses the computed Binomial likelihoods for the data observed to update the prior distributions on the Beta to compute posterior distributions for all mortality probability parameters. Full details and results are given in the Appendix. For further background information on this type of Bayesian analysis see (Fenton & Neil, 2018).

The four hypotheses being tested (denoted $H1 - H4$) about the causal connections between variables deaths ($D$), Covid-19 Severity ($S$), and Treatment ($T$), are as follows:

$H1$: $P(D)$ – death is independent of Covid-19 severity or treatment

$H2$: $P(D|S)$ – death is dependent on Covid-19 severity only

$H3$: $P(D|T)$ – death is dependent on treatment only

$H4$: $P(D|S,T)$ – death is dependent on Covid-19 severity and treatment

These hypotheses are shown graphically in Figure 1.

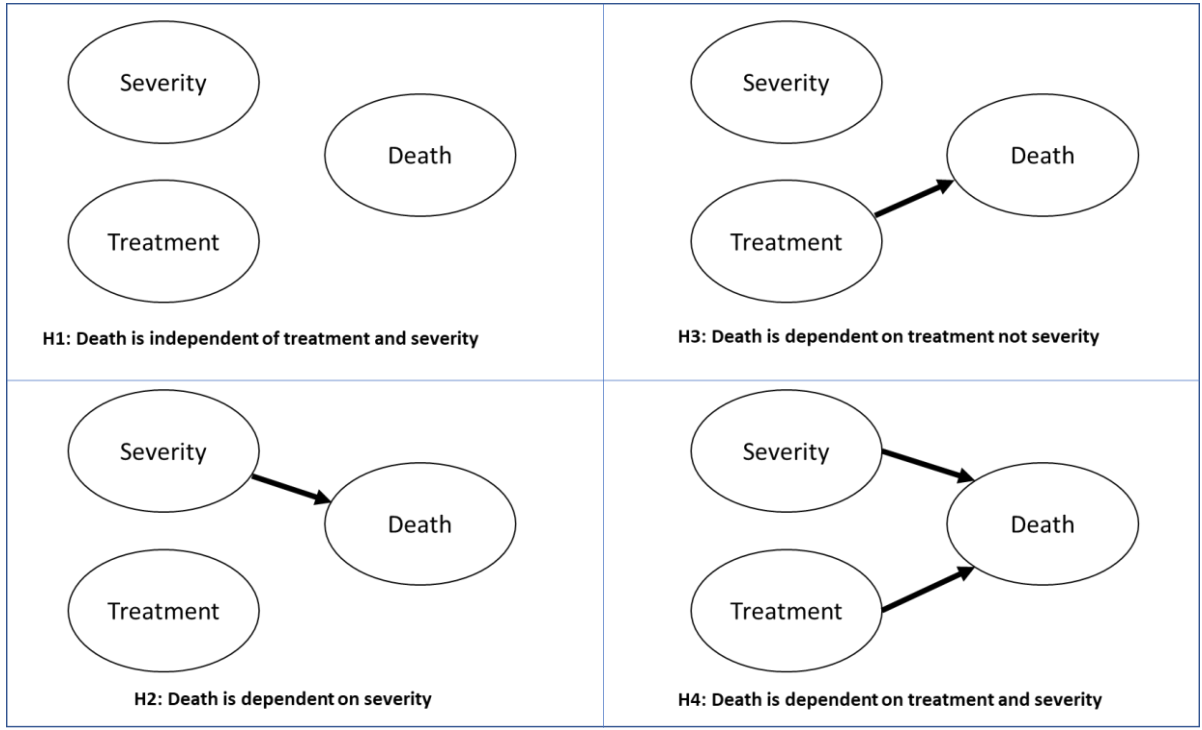

Figure 1: Causal hypotheses $H1, H2, H3, H4$

From applying the analysis stages, A to D, the resulting posterior probability of these hypotheses being true given the data is:

$$P(H1\,|Data) = 0, P(H2\,|Data) = 0.0092, P(H3\,|Data) = 0, P(H4\,|Data) = 0.9908$$

Hence, there is extremely convincing evidence that Covid-19 severity and treatment causally influence mortality.

To estimate the magnitude of the impact of Covid-19 severity, $S$, and Treatment, $T$, on death, $D$ we need to compute $H4: P(D|S,T)$. Figure 2 shows the marginal probability distributions for mortality for each of the combinations of severity and treatment. While we can still compute the mean and confidence intervals (CIs) for these distributions (as shown in Table 2), in contrast to the classical approach these CIs do not rely on notions of repeated trials. Also, the classical CIs hide crucial information about where the probability mass is located.

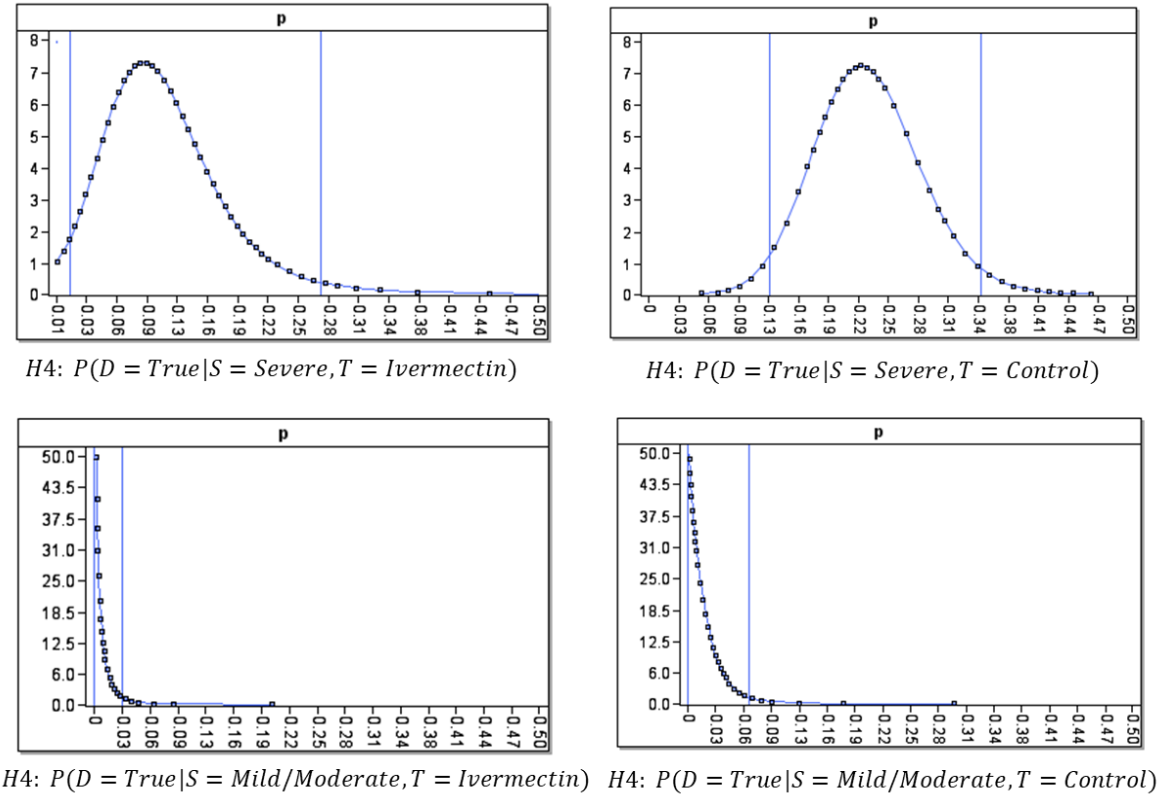

H4: $P(D = True|S = Severe, T = Ivermectin)$  H4: $P(D = True|S = Severe, T = Control)$

H4: $P(D = True|S = Mild/Moderate, T = Ivermectin)$  H4: $P(D = True|S = Mild/Moderate, T = Control)$

Figure 2: Posterior marginal probability distributions for mortality $p$ from meta-analysis

|  | Median | Mean | 95% CI |
|---|---|---|---|
| $P(D = True|S = Severe, T = Ivermectin)$ | 0.107 | 0.117 | (0.019, 0.275) |
| $P(D = True|S = Severe, T = Control)$ | 0.227 | 0.229 | (0.125, 0.349) |
| $P(D = True|S = Mild/Moderate, T = Ivermectin)$ | 0.0003 | 0.004 | (0, 0.0036) |
| $P(D = True|S = Mild/Moderate, T = Control)$ | 0.012 | 0.0178 | 0, 0.068) |

Table 2: Mean and 95% confidence intervals.

The risk ratio $RR$ is the estimated mortality probability of ivermectin patients divided by the estimated mortality probability of control patients. One of the advantages of the Bayesian approach is that the shape and scale of the probability distribution for $RR$ can be directly calculated and inspected whilst making minimal statistical assumptions. Figure 3 shows the marginal probability distribution of $RR$. Note that the probability distribution $RR$ for mild/moderate Covid-19 is heavily asymmetric because the lower bounds for $P(D = True)$ are zero (see Table 2), hence producing a zero-division computational overflow. For this reason, classical statistical methods cannot easily estimate this quantity. However, we can instead use an arithmetically alternative measure that does not suffer from this defect, risk difference, $RD = ivermectin - control$. The marginal probability distribution for $RD$ is also shown in Figure 3. For mild/moderate Covid-19 there is a clear modal spike around zero, though with most of the probability mass of this notably skewed distribution lying in the region $RD < 0$, favouring ivermectin. The centroid of the probability mass is closer to zero difference than for severe Covid-19, where the centroid is further away from zero. This indicates a stronger effect of ivermectin treatment on mortality in severe disease than for mild/moderate. It also suggests our confidence in the evidence for ivermectin treatment for severe Covid-19 is stronger than for mild/moderate Covid-19, though this is a quantitative question based on the probability mass for $RR < 1$ or $RD < 0$.

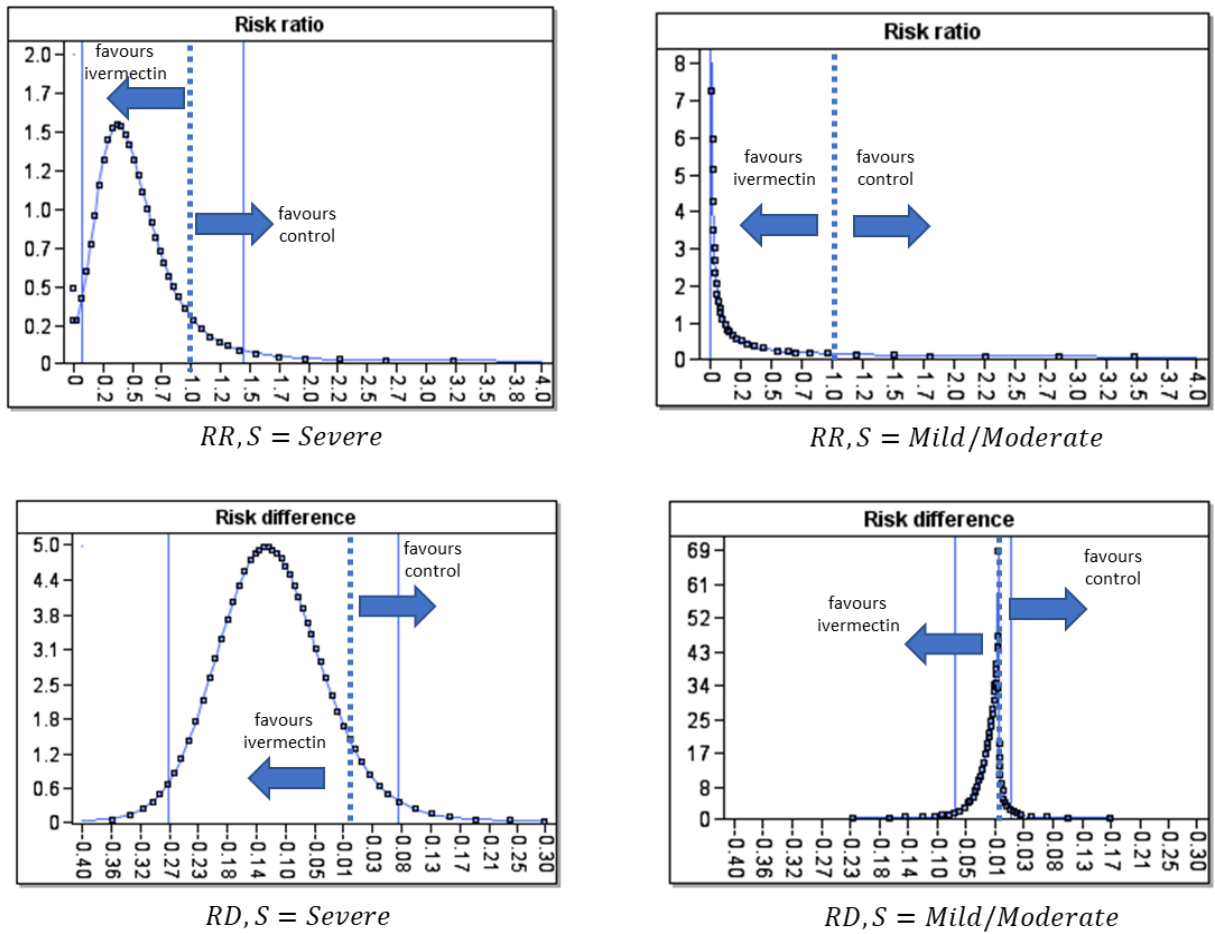

Figure 3: Posterior marginal probability distributions for $RR$ and $RD$ from meta-analysis

If the $RR$ is less than one, then this provides support for the hypothesis that the treatment is effective (the lower the number the more effective) and if the upper bound of the confidence interval for the $RR$ is less than one then it is conventionally concluded that the treatment is effective with that level of confidence (95% in this case). In the Bayesian approach, from the marginal probability distributions shown in Figure 3, we compute the probability that the risk ratio, $RR < 1$, dependent on the severity of Covid-19, as shown in Table 3.

|  | Severe | Mild to Moderate |
| --- | --- | --- |
| $P(RR < 1)$ | 90.7% | 84.1% |

Table 3: Probability of risk ratio, $RR < 1$, favouring ivermectin vs control

The $RR$ results of the previous studies, together with the $RR$ results from our Bayesian analysis, are shown in Table 4.

|  | *RR* | *RR* 95% CI |
|---|---|---|
| WHO 2021 | 0.19 | (0.09, 0.38) |
| Roman et al, 2021 (all mild or moderate cases) | 0.37 | (0.12, 1.13) |
| Popp et al, 2021 (all mild or moderate cases) | 0.60 | (0.14, 2.51) |
| Bryant et al, 2021 (mild or moderate cases) | 0.24 | (0.06, 0.94) |
| Bryant et al, 2021 (severe cases) | 0.51 | (0.22, 1.14) |
| Bryant et al, 2021 (all cases) | 0.38 | (0.19, 0.73) |
| Bayesian analysis, 2021 (mild or moderate cases) | 0.34 | (0.00, 26.0) |
| Bayesian analysis, 2021 (severe cases) | 0.48 | (0.08, 1.46) |

Table 4: Summary of risk ratio results from previous studies and this Bayesian analysis

The $RD$ results from our Bayesian analysis are shown in Table 5.

|  | *RD* | *RD* 95% CI |
|---|---|---|
| Bayesian analysis, 2021 (mild or moderate cases) | -0.013 | (-0.066, 0.020) |
| Bayesian analysis, 2021 (severe cases) | -0.110 | (-0.269, 0.076) |

Table 5: Summary of risk difference results from this Bayesian meta-analysis

## 4. Sensitivity Analysis

We perform a sensitivity analysis to determine the extent to which the results depend on the trial data from a particular study. We do this by removing one study at a time and reformulating the model without that data set. The sensitivity analysis on the risk ratio and difference results are shown in Table 7.

|  | Risk Ratio ($RR$) | | Risk Difference ($RD$) | | $P(RD < 0)$ |
|---|---|---|---|---|---|
|  | Median | 95% CI | Mean | 95% CI | $P(RR < 1)$ |
| Mild to Moderate |  |  |  |  |  |
| Ahmed 2020 | 0.03 | (0,22) | -0.014 | (-0.06,0.02) | 0.84 |
| Babalola 2020 | 0.03 | (0,23) | -0.014 | (-0.06,0.02) | 0.84 |
| Chaccour 2020 | 0.03 | (0,24) | -0.014 | (-0.07,0.02) | 0.84 |
| Elgazzar 2020 | 0.07 | (0,156) | -0.009 | (-0.06,0.02) | 0.78 |
| Hashim 2020 | 0.03 | (0,17) | -0.014 | (-0.07,0.02) | 0.85 |
| Lopez-Medina 2021 | 0.05 | (0,36) | -0.015 | (-0.07,0.02) | 0.83 |
| Mahmud 2020 | 0.06 | (0,135) | -0.011 | (-0.07,0.02) | 0.79 |
| Mohan 2021 | 0.04 | (0,18) | -0.015 | (-0.07,0.02) | 0.85 |
| Petkov 2021 | 0.03 | (0,17) | -0.015 | (-0.07,0.02) | 0.85 |
| Ravikirti 2021 | 0.06 | (0,145) | -0.005 | (-0.05,0.02) | 0.78 |
| Rezai 2020 | 2.E-04 | (0,8) | -0.017 | (-0.07,0.01) | 0.91 |
| All included | 0.03 | (0,26) | -0.013 | (-0.07,0.02) | 0.84 |
| Severe |  |  |  |  |  |
| Elgazzar 2020 | 0.72 | (0.24,1.73) | -0.07 | (-0.23,0.11) | 0.77 |
| Fonseca 2021 | 0.34 | (0.05,1.16) | -0.15 | (-0.30,0.02) | 0.96 |
| Gonzalez 2021 | 0.41 | (0.04,1.43) | -0.13 | (-0.29,0.07) | 0.92 |
| Hashim 2020 | 0.63 | (0.09,1.75) | -0.09 | (-0.26,0.12) | 0.86 |
| Okumus 2021 | 0.43 | (0.04,1.54) | -0.11 | (-0.27,0.08) | 0.90 |
| All included | 0.48 | (0.08,1.46) | -0.11 | (-0.27,0.08) | 0.91 |

Table 7: Risk ratio and difference summary statistics for each study removed one at a time

For Mild to Moderate Covid-19 the removal of any single study clearly does not substantially affect the conclusion, with $P(RD < 0)$ lying in the range {0.78, 0.91}. For Severe Covid-19 this range is {0.77, 0.96} suggesting that if the (Fonseca 2021) or (Gonzalez 2021) were removed from the meta-analysis the support for the effectiveness of the treatment would substantially improve, and in the case of removing (Fonseca 2021) this would increase confidence, that the risk difference is below zero, beyond 95%.

We can re-examine the causality hypothesis under the most unfavourable conditions to the difference hypothesis, which occurs when the (Elgazzar 2020) trial is removed. The posterior probability distribution for the four hypotheses are:

$$P(H1|Data) = 0, P(H2|Data) = 0.088, P(H3|Data) = 0, P(H4|Data) = 0.9118$$

So, clearly the causality hypothesis test still strongly supports $H4$ even in the absence of (Elgazzar 2020).

## 5. Conclusions

This Bayesian meta-analysis has shown that there is a 99% posterior probability for the hypothesis that mortality is causally dependent on both Covid-19 severity and ivermectin treatment. The Bayesian meta-analysis estimates that the mean probability of death of patients with severe Covid-19 is 11.7% (CI 1.9 – 27.5%) for those given ivermectin, compared to 22.9% (CI 12.5 – 34.9%) for those not given ivermectin. For the severe Covid-19 cases the probability of the risk ratio being less than one is 90.7% while for mild/moderate cases this probability it is 84.1%.

By removing one study at a time, we were able to evaluate the sensitivity of the conclusions to a single study. In the worst case, where (Elgazzar 2020) is removed the results remain robust, for both severe and mild to moderate Covid-19. It should be noted that the composite study of (Niaee 2021) was already excluded for the reasons given. Also, we can identify those studies, which, were they *not* to be included would lead to an *increase* in the confidence in the treatment effect.

In our view this Bayesian analysis, based on the statistical study of the RCT data, provides sufficient confidence that ivermectin is an effective treatment for Covid-19 in reducing mortality. This belief supports the conclusions of (Bryant et al., 2021) over those of (Roman et al., 2021). The conclusions of (Roman et al., 2021) are based on the subjective assessment that the RCTs were 'low quality' but even taking this into account simply means weaker evidence of a positive effect, rather than 'no effect'. Moreover, it is important to point out that there are also many observational studies which provide additional evidence of the effectiveness of ivermectin (CovidAnalysis, 2021; Kory et al., 2021; Santin, Scheim, McCullough, Yagisawa, & Borody, 2021). Unlike our analysis, which was restricted to effect on mortality, this includes evidence of the effectiveness of ivermectin in reducing infection or hospitalizations when taken early.

The paper has also highlighted the advantages of using Bayesian methods over classical statistical methods for meta-analysis, which is especially persuasive in providing a transparent marginal probability distribution for both risk ratio $RR$ and risk difference, $RD$. Furthermore, we show that using $RD$ avoids the estimation and computational issues encountered using $RR$, thus making full and more efficient use of all evidence, without *ad hoc* "continuity corrections" for avoidance of division-by-zero anomalies.

**Data and Models**

All of the models and data used in this work are available in a zip file, which can be downloaded from: http://www.eecs.qmul.ac.uk/~norman/Models/ivermectin_models.zip

The models can all be run using the free trial version of AgenaRisk:
https://www.agenarisk.com/agenarisk-free-trial

## References

Agena Ltd. (2021). *AgenaRisk*. Retrieved from http://www.agenarisk.com

Bryant, A., Lawrie, T. A., Dowswell, T., Fordham, E. J., Mitchell, S., Hill, S. R., & Tham, T. C. (2021). Ivermectin for Prevention and Treatment of COVID-19 Infection: A Systematic Review, Meta-analysis, and Trial Sequential Analysis to Inform Clinical Guidelines. *American Journal of Therapeutics*. https://doi.org/10.1097/MJT.0000000000001402

CovidAnalysis. (2021). Ivermectin for COVID-19: real-time meta analysis of 63 studies. Retrieved from https://c19ivermectin.com/

Crawford, M. (2021). The Meta-Analytical Fixers: An Ivermectin Tale. Retrieved July 7, 2021, from https://roundingtheearth.substack.com/p/the-meta-analytical-fixers-an-ivermectin

Elgazzar, A., Hany, B., Youssef, S. A., Hafez, M., Moussa, H., & Eltaweel, A. (2020). *Efficacy and Safety of Ivermectin for Treatment and prophylaxis of COVID-19 Pandemic.* https://doi.org/10.21203/RS.3.RS-100956/V2

Fenton, N. E., & Neil, M. (2018). *Risk Assessment and Decision Analysis with Bayesian Networks* (2nd ed.). CRC Press, Boca Raton.

Fordham, E. J., & Lawrie, T. A. (2021). Attempt to discredit landmark British ivermectin study. Retrieved July 7, 2021, from HART Health Advisory & Recovery Team website: https://www.hartgroup.org/bbc-ivermectin/

Gelman, A., Carlin, J. B., Stern, H. S., Dunson, D. B., Vehtari, A., & Rubin, D. B. (2013). *Bayesian Data Analysis*. https://doi.org/10.1201/b16018

Kory, P., Meduri, G. U., Varon, J., Iglesias, J., & Marik, P. E. (2021). Review of the Emerging Evidence Demonstrating the Efficacy of Ivermectin in the Prophylaxis and Treatment of COVID-19. *American Journal of Therapeutics*, *28*(3), e299–e318. https://doi.org/10.1097/MJT.0000000000001377

Neil, M., & Fenton, N. (2021). Bayesian Hypothesis Testing and Hierarchical Modeling of Ivermectin Effectiveness. *American Journal of Therapeutics*, *28*(5), e576–e579. https://doi.org/10.1097/MJT.0000000000001450

Niaee, M. S., Namdar, P., Allami, A., Zolghadr, L., Javadi, A., Karampour, A., … Gheibi, N. (2021). Ivermectin as an adjunct treatment for hospitalized adult COVID-19 patients: A randomized multi-center clinical trial. *Asian Pacific Journal of Tropical Medicine*, *14*(6), 266. https://doi.org/10.4103/1995-7645.318304

Popp, M., Stegemann, M., Metzendorf, M.-I., Gould, S., Kranke, P., Meybohm, P., … Weibel, S. (2021). Ivermectin for preventing and treating COVID-19. *Cochrane Database of Systematic Reviews*, *2021*(8). https://doi.org/10.1002/14651858.CD015017.pub2

Roman, Y. M., Burela, P. A., Pasupuleti, V., Piscoya, A., Vidal, J. E., & Hernandez, A. V. (2021). Ivermectin for the treatment of COVID-19: A systematic review and meta-analysis of randomized controlled trials. *MedRxiv*, 2021.05.21.21257595. https://doi.org/10.1101/2021.05.21.21257595

Santin, A. D., Scheim, D. E., McCullough, P. A., Yagisawa, M., & Borody, T. J. (2021). Ivermectin: a multifaceted drug of Nobel prize-honored distinction with indicated efficacy against a new global scourge, COVID-19. *New Microbes and New Infections*, 100924. https://doi.org/10.1016/J.NMNI.2021.100924

Sutton, A. J., & Abrams, K. R. (2001). Bayesian methods in meta-analysis and evidence synthesis. *Statistical Methods in Medical Research*, *10*(4), 277–303.

WHO (World Health organization). (2021). *Therapeutics and COVID-19: Living Guideline*. Retrieved from https://www.who.int/publications/i/item/WHO-2019-nCoV-therapeutics-2021.2

## Appendix

The stages in the analysis are organised as follows:

A. Learn the mortality probability distribution from relevant trials for each hypothesis of concern using a Beta-Binomial hierarchical model.
B. For each causal hypothesis use the model in stage A to learn the mortality probability distributions relevant to that causal hypothesis.
C. For each causal hypothesis use the learnt probability distributions from stage B to predict the observed data and calculate the likelihood of observing that data.
D. For all causal hypotheses compute the posterior probability of each hypothesis given the likelihood of observing the data under that hypothesis and select the most likely causal hypothesis that explains the data.
E. Estimate the magnitude of impact of the relevant variables, under that hypothesis, on mortality.

For each hypothesis and combination of Covid-19 severity and treatment variable state we learn the corresponding mortality probability distribution using a hierarchical Beta-Binomial model (where $m$ is the number of studies, $n_i$ is the number of patients and $x_i$ is the number of deaths in study $i$):

$$P(p) = \sum_{\alpha,\beta,p_i,x_i} P(p|\alpha,\beta) P(\alpha,\beta) \left\{ \prod_{i=0}^{m} P(x_i|n_i,p_i) P(p_i|\alpha,\beta) \right\}$$

$$P(x_i|n_i,p_i) = \binom{n_i}{x_i} p_i^{x_i} (1-p_i)^{n_i-x_i}$$

$$p_i \sim Beta(\alpha,\beta)$$

$$\alpha,\beta \sim Uniform(0,100)$$

where the mortality probability, $p$, is determined by two parameters, $\alpha$ and $\beta$ that model the global distribution of $p_i$ variables across the studies, where each $p_i$ is determined by its local data $(n_i, x_i)$. The prior distributions chosen for $\alpha, \beta$ induce an ignorant prior with a mean $p = 0.5$ with a distribution broadly flat in the range [0, 1]. Note that here the same prior is used for $p$ and $p_i$ thus favouring neither control nor treatment.

An example of the structure of the Bayesian model used in steps A to C is shown in Figure 4, as a Bayesian Network, where we learn the probability distribution for $p \equiv P(D = True | S = Severe, T = Ivermectin)$ from the $m = 5$ relevant studies using data pairs $(n_i, x_i)$ for deaths and number of subjects in given trial.

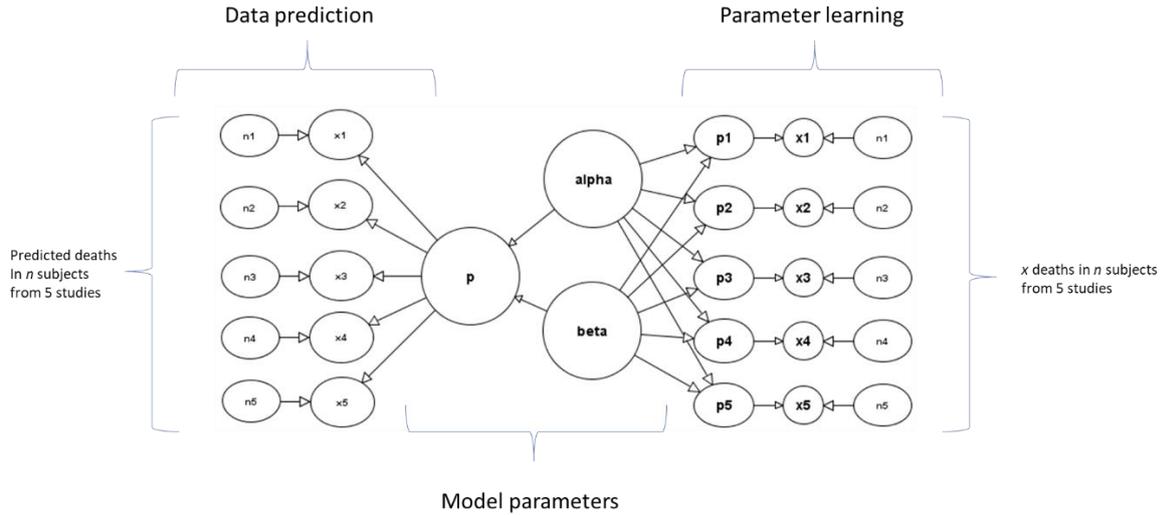

Figure 4: Meta analysis Bayesian Network

Once we have learnt $P(p|\alpha,\beta)$ from the data we need to determine how well the learnt distribution explains that data under each hypothesis $H1 - H4$. Note that each hypothesis has a different number of mortality probability parameters, $p$, determined by the number of states for each variable for that hypothesis. So, for $H1: P(D)$ we only have one probability to determine. For $H2: P(D|S)$ we have two mortality probabilities to consider, one for severe Covid-19 and another for mild-moderate Covid-19, and so on.

As the number of mortality probability parameters to be estimated under each hypothesis increases the smaller the amount of data available to estimate each one. This leads to greater variance in predictions of the data when there are more parameters and, thus, models with more parameters are penalised by Occam's razor.

To test the predictions of the data under each hypothesis, $H_i$, we use Bayes:

$$P(H_i|Data) \propto P(Data|H_i)P(H_i)$$

Here we assume the prior probabilities $P(H_i)$ are uniform and we can calculate $P(Data|H_i)$ as:

$$P(Data|H_i) = \prod_{i=0}^{m} P(x_i|n_i, p_{H_i})$$

which is simply the product of likelihoods over all trials data, using the learnt $p_{H_i}$ variables for the given hypothesis. Given the uniform prior assumption the posterior belief in each causal hypothesis is simply: $P(H_i|Data) = P(Data|H_i)$. The results are shown in Table 5.

| Hypothesis | Summary statistics for learnt $p$ distributions | | | | Likelihood of Data given $p$ | | Joint | Posterior |
|---|---|---|---|---|---|---|---|---|
| | | Median | Mean | 95% CI | | Likelihood | likelihood | probability |
| H1 | P(Death) | 1.11% | 5.78% | (0, 35.8) | P(Data) | 2.97E-28 | 2.97E-28 | 0.000 |
| H2 | P(Death \| C = Severe) | 16.52% | 17.20% | (5.5, 33.13) | P(Data \| C = Severe) | 5.65E-13 | 1.29E-21 | 0.009 |
| | P(Death \| C = Mild/Moderate) | 0.31% | 0.86% | (0, 4.74) | P(Data \| C = Mild/Moderate) | 2.29E-09 | | |
| H3 | P(Death \| T = Ivermectin) | 0.04% | 3.37% | (0, 23.35) | P(Data \| T = Ivermectin) | 4.30E-11 | 6.86E-28 | 0.000 |
| | P(Death \| T = Control) | 3.63% | 7.62% | (0, 37.82) | P(Data \| T = Control) | 1.60E-17 | | |
| H4 | P(Death \| S = Severe, T = Ivermectin) | 10.74% | 11.71% | (1.93, 27.62) | P(Data \| S = Severe, T = Ivermectin) | 2.17E-06 | 1.40E-19 | 0.991 |
| | P(Death \| S = Mild/Moderate, T = Ivermectin) | 0.03% | 0.42% | (0, 3.13) | P(Data \| S = Mild/Moderate, T = Ivermectin) | 1.24E-02 | | |
| | P(Death \| S = Severe, T = Control) | 22.65% | 22.91% | (12,6, 34.75) | P(Data \| S = Severe, T = Control) | 3.95E-06 | | |
| | P(Death \| S = Mild/Moderate, T =Control) | 1.20% | 1.78% | (0, 6.89) | P(Data \| S = Mild/Moderate, T = Control) | 1.32E-06 | | |

Table 5: Summary statistics of distributions and resulting likelihood predictions

The above description takes us up to stage D and established the support for each causal hypothesis. Here there was overwhelming support for hypothesis $H4$ and hence we use the causal structure for this hypothesis to compute the necessary impact statistics at stage E:

- compute the risk ratio ($RR$).
- compute the risk difference ($RD$).
- determine the probability of the risk ratio being less than one.

The relevant computations here are:

$$RR = \frac{p_{ivermectin}}{p_{control}}$$

$$RD = p_{ivermectin} - p_{control}$$

$$P(RR < 1) = \int_0^1 f(RR)\,dRR$$

All calculations are carried out using AgenaRisk Bayesian network software (Agena Ltd, 2021). All models used are available on request and all can be run in the free trial version of AgenaRisk.